# A new expression of the probability distribution in incomplete statistics and fundamental thermodynamic relations

ZhiFu Huang[1], Bihong Lin[1,2] and Jincan Chen[1,*]

[1] Department of Physics, Xiamen University, Xiamen 361005, People's Republic of China

[2] Department of Physics, Quanzhou Normal University, Quanzhou 362000, People's Republic of China

**Abstract**

In order to overcome the limitations of the original expression of the probability distribution appearing in literature of Incomplete Statistics, a new expression of the probability distribution is derived, where the Lagrange multiplier $\beta$ introduced here is proved to be identical with that introduced in the second and third choices for the internal energy constraint in Tsallis' statistics and to be just equal to the physical inverse temperature. It is expounded that the probability distribution described by the new expression is invariant through uniform translation of the energy spectrum. Moreover, several fundamental thermodynamic relations are given and the relationship between the new and the original expressions of the probability distribution is discussed.

PACS numbers: 05.20.-y; 05.30.-d; 05.70.-a; 02.50.-r

---

Email: jcchen@xmu.edu.cn

# 1. Introduction

Like Tsallis' statistics [1-4], Incomplete Statistics (IS) proposed by Wang [5] in 2001 and developed by some researchers [6-18] has been an important part of Nonextensive Statistical Mechanics. It may be used to investigate the thermostatistic properties of the physical systems with fractal and self-similar structures, long-range interacting and/or long-duration memory, anomalous diffusion phenomena, and so on. Although a lot of important conclusions have been obtained, some fundamental problems have not been solved yet. For example, the Lagrange multiplier $^w\beta$ introduced by Wang [5] and adopted in Refs.[6, 8, 9, 13, 15, 17] can not be defined as the physical inverse temperature, and consequently, a concomitant definition of the physical temperature has to be given [19]. And the probability distribution derived by Wang [5] and adopted by other researchers [6-11, 13-19] varies with the uniform translation of the energy spectrum, and consequently, the probability distribution will depend on the choice of the zero point of internal energy. Thus, the theory of IS needs to be further developed and improved.

In the present paper, the entropy of IS is directly used to derive a new expression of the probability distribution, from which the limitations of the original expression of the probability distribution are overcome and some significant results are obtained.

# 2. Limitations of the original probability distribution in IS

On the basis of IS proposed by Wang [5], the q-entropy for a nonextensive system is given by

$$S_q = k \frac{\sum_{i=1}^{w} p_i - 1}{q-1} \qquad (1)$$

with the incomplete normalization

$$\sum_{i}^{w} p_i^q = 1 \qquad (2)$$

and the internal energy of the system is determined by

$$U_q = \sum_i^w p_i^q \varepsilon_i, \tag{3}$$

where $k$ is the Boltzmann constant and taken as 1 hereafter, $p_i$ is the probability of the state $i$ among $W$ possible ones that are accessible to the calculation, $\varepsilon_i$ is the energy of the system at state $i$, and q is the nonextensive parameter. For the sake of convenience, $\sum_{i=1}^w$ is replaced by $\sum_i$ below.

Using Eqs. (1)-(3) and the Lagrange equation [5]

$$\delta(S_q + \frac{{}^w\alpha}{1-q}\sum_i^w p_i^q - {}^w\alpha\, {}^w\beta U_q) = 0, \tag{4}$$

Wang [5] first derived the probability distribution of IS as

$$p_i = \frac{[1-(1-q)\,{}^w\beta \varepsilon_i]^{1/(1-q)}}{{}^wZ_q} \tag{5}$$

with

$${}^wZ_q = \{\sum_i [1-(1-q)\,{}^w\beta \varepsilon_i]^{q/(1-q)}\}^{1/q}, \tag{6}$$

where ${}^w\alpha$ and ${}^w\beta$ are the Lagrange multipliers introduced in the original paper of IS [5]. It is worthwhile pointing out that ${}^w\beta$ in IS [5, 6, 8, 9,13, 15, 17] is not the inverse temperature, while the Lagrange multiplier ${}^T\beta$ introduced in the second and third choices for the internal energy constraint in Tsallis' statistics is the inverse temperature.

On the basis of the above results, it has been strictly proven that [19]

$$\frac{\partial S_q}{\partial U_q} = \frac{{}^w\beta\, {}^wZ_q^{q-1}}{q} = \beta' = \frac{1}{T}, \tag{7}$$

where $T$ is the physical temperature of the system in equilibrium and $\beta'$ is the deformed Lagrange multiplier, which is identical with the Lagrange multiplier ${}^T\beta$ in Tsallis' statistics.

On the other hand, it is clearly seen from Eq. (5) that if $\varepsilon_i$ is replaced by $\varepsilon_i + \varepsilon_0$, the form of the probability distribution will be changed, i.e.,

$$p_i(\varepsilon_i) \neq p_i(\varepsilon_i + \varepsilon_0), \tag{8}$$

where $\varepsilon_0$ is a constant. It implies that the probability distribution function depends on the choice of the zero point of internal energy.

### 3. A new expression of the probability distribution

By using Eqs. (1)-(3), the Lagrange method can be written as

$$\delta(S_q - \alpha \sum_i p_i^q - \beta U_q) = 0, \tag{9}$$

where $\alpha$ and $\beta$ are the Lagrange multipliers. From Eqs. (1)-(4), we obtains

$$\frac{1}{q-1} - \alpha q p_i^{q-1} - \beta q p_i^{q-1} \varepsilon_i = 0, \tag{10}$$

$$\alpha = \frac{1}{q}\left(\frac{\sum_i p_i}{q-1} - q\beta U_q\right), \tag{11}$$

and

$$p_i = [\sum_{j=1}^w p_j - (1-q)q\beta(\varepsilon_i - U_q)]^{\frac{1}{1-q}}. \tag{12}$$

Equation (12) may be written as another form:

$$p_i = \frac{1}{Z_q}[1 - (1-q)q\beta(\varepsilon_i - U_q)/\sum_{j=1}^w p_j]^{\frac{1}{1-q}} \tag{13}$$

with

$$Z_q = \left(\sum_{i=1}^w p_i\right)^{\frac{1}{q-1}} = [\sum_{i=1}^w (1-(1-q)q\beta(\varepsilon_i - U_q)/\sum_{j=1}^w p_j)^{\frac{q}{1-q}}]^{\frac{1}{q}}. \tag{14}$$

Equation (13) is a new expression of the probability distribution function and can overcome the limitations existing in the original distribution function.

It can be proved by using Eq. (12) that

$$\frac{\partial \sum_j p_j}{\partial \beta} = \frac{1}{1-q} \sum_i \left\{ p_i^q [\frac{\partial \sum_j p_j}{\partial \beta} - (1-q)q(\varepsilon_i - U_q) + (1-q)q\beta \frac{\partial U_q}{\partial \beta}] \right\}$$

$$= \frac{1}{1-q} \frac{\partial \sum_j p_j}{\partial \beta} + q\beta \frac{\partial U_q}{\partial \beta}, \qquad (15)$$

and consequently,

$$\frac{\partial S_q}{\partial U_q} = \beta = \beta' \qquad (16)$$

It shows clearly that Lagrange multiplier $\beta$ introduced here is identical with the deformed Lagrange multiplier $\beta'$ defined in Ref. [19]. Like Tsallis' statistics, the Lagrange multiplier $\beta$ appearing in the new expression of the probability distribution function is just the physical inverse temperature. As described in Ref. [20], the Legendre transform structure of thermodynamics does not depend upon the functional form of the entropy. The conclusion conforms to Abe's standpoint [21], i.e., statistical mechanics may be modified but thermodynamics should remain unchanged.

It is clearly seen from Eqs. (3) and (13) that if $\varepsilon_i$ is replaced by $\varepsilon_i + \varepsilon_0$, $U_q$ becomes $U_q + \varepsilon_0$, but $\varepsilon_i - U_q$ is invariant so that the probability distribution function is invariant through uniform translation $\varepsilon_0$ of the energy spectrum $\{\varepsilon_i\}$.

### 4. Several fundamental thermodynamic relations

Using Eqs. (1) and (14), one may define a new parameter $\bar{Z}_q$ as

$$\ln_q \bar{Z}_q = S_q - \beta U_q = Z_q^{q-1} \ln_q Z_q - \beta U_q, \qquad (17)$$

where $\ln_q Z_q = (Z_q^{1-q} - 1)/(1-q)$ is the generalized logarithm function [19, 22, 23]. From Eqs. (16) and (17), we can derive several important thermodynamic relations as

$$U_q = -\frac{\partial}{\partial \beta} \ln_q \bar{Z}_q, \qquad (18)$$

$$F_q = U_q - TS_q = -\frac{1}{\beta}\ln_q \overline{Z}_q, \qquad (19)$$

and

$$S_q = -\frac{\partial F_q}{\partial T}, \qquad (20)$$

where $F_q$ is the free energy of the system. It is seen from the above results that the forms of Eqs. (18)-(20) are the same as those in Boltzmann-Gibbs (BG) statistics. When $q \to 1$, the above results reduce to the fundamental thermodynamic relations in BG statistics.

**5. Discussion**

Comparing Eq. (4) and Eq. (9), one can easily find that the relations between the different Lagrange multipliers $^w\alpha$, $^w\beta$, $\alpha$ and $\beta$ are, respectively, determined by

$$\frac{^w\alpha}{1-q} = -\alpha \qquad (21)$$

and

$$^w\alpha\, ^w\beta = \beta. \qquad (22)$$

Using Eqs. (11), (21) and (22), we obtain

$$^w\alpha = \sum_i p_i / q + (1-q)\beta U_q \qquad (23)$$

and

$$^w\beta = \frac{q\beta}{\sum_i p_i + (1-q)q\beta U_q}. \qquad (24)$$

By substituting Eq. (24) into Eq. (4), it is very natural to obtain Eq. (13) and the following relation

$$^wZ_q^{q-1} = Z_q^{q-1}[1 + (1-q)q\beta U_q / \sum_i p_i] = \sum_i p_i + (1-q)q\beta U_q. \qquad (25)$$

It shows clearly that although the new expression of the distribution function is equivalent with the original expression of the distribution function in IS, it can overcome the limitations existing in the original distribution function. Thus, it may be expected that the results obtained here will be helpful for the further development and improvement of IS.

**Acknowledgments**

This work has been supported by the Research Foundation of Ministry of Education, People's Republic of China.